\documentclass[a4paper,11pt]{article}
\pdfoutput=1 % if your are submitting a pdflatex (i.e. if you have
\usepackage{jheppub} % for details on the use of the package, please
\usepackage[T1]{fontenc} % if needed

\usepackage[utf8]{inputenc}
\usepackage[T1]{fontenc}

\title{\boldmath Physical observers, $T$-vacuum and Unruh like effect in the radiation dominated early universe}

\author[a,b,c]{Sujoy K. Modak}

\affiliation[a]{Facultad de Ciencias - CUICBAS, Universidad de Colima, \\Colima, C.P. 28045, M\'exico}
\affiliation[b]{KEK Theory Center, High Energy Accelerator Research Organization (KEK),\\ Tsukuba, Ibaraki 305-0801, Japan}
\affiliation[c]{Physics Department, California State University,\\ Fresno, CA 93740-8031, USA}

\emailAdd{smodak@ucol.mx}

\abstract
{We report the existence of an Unruh like effect, for physical observers (cosmological and comoving observers) in the radiation dominated early universe, which is possible due to the discovery of {\em a new vacuum state} (referred here as the $T-$vacuum). Both the comoving and the cosmological observers, who are crucial in our understanding of cosmology, observe this $T-$vacuum as a particle excited state and are able to detect radiation due to particle creation. We draw a robust analogy with the Unruh effect, whereby -- (i) the physical observers here are closely analogous to the accelerated (Rindler) observers in the flat spacetime, and (ii) the $T-$vacuum  plays the role of the Minkowski vacuum state which contains particles when viewed from the physical observers frame. Our analogy is further supported by a proof of well-defined (hadamard) behavior of the $T-$vacuum in the entire spacetime. All our analysis of the particle creation process is done here within a two dimensional set up.}
\begin{document} 
\maketitle
%\flushbottom

\section{Introduction}
In general relativity, although coordinate systems and observers, by themselves, do not have a role to dictate fundamental physical laws (which should be covariantly defined), the choice of one frame over the other often helps to gain important physical insights. We see this happening in black holes where one has options to chose coordinates so that the coordinate singularity at the event horizon may appear (e.g.,  Schwarzschild coordinates) or may not appear (e.g., Kruskal coordinates). Also, if we want to single out a physical observer (like the asymptotic observer in Schwarzschild spacetime) we need to select a coordinate system suitable to that particular observer (i.e., Schwarzschild coordinates) in order to discuss relevant physics. This, to some extent is also true even in a flat spacetime where an observer with constant four acceleration encounters a horizon in Minkowski spacetime (e.g. Rindler observer) just because Rindler coordinates only cover one-fourth of the Minkowski spacetime. Physics become even more interesting by including quantum fields (even non-interacting case) into account which then introduces particle creation due to gravitational field (e.g., Hawking effect \cite{hawk1, hawk2}), due to the observer's own motion (e.g., Unruh effect \cite{unruh}) etc. However, detecting any radiation coming from these particle creation is quite unlikely because of the following reasons -- (i) for the Hawking effect, black hole temperature is inversely proportional to its mass and this makes the radiation quite faint and hard to detect for astrophysical black holes with larger masses. On the top of that there is  Cosmic Microwave Background (CMB) radiation which provides a uniform background temperature of nearly 2.7 Kelvin and anything below this is hard to detect. And, (ii) for the Unruh effect,  the particle detector itself is meant to be carried along with the accelerated observer and this acceleration needs to be very large, far beyond of what we can achieve in a laboratory setting, for any direct detection. Thus hope for a direct detection of one/both of these effects, which are although quite straight-forward from theoretical considerations, is/are quite less and unlikely to happen in the near future. Nevertheless,  there are ongoing efforts to verify these effects in the analogue gravity systems which could be indicative but also suffer from the limitation of being tagged as a ``toy model''.

In this paper we report a new example of particle creation in the cosmological setting,  particularly in the context of the radiation dominated phase of the early universe. This is an extension of our earlier work \cite{modak1} and carries a decisive advancement of the observations carried out in \cite{modak1}. We show here, that physical observers, such as given by the ``cosmological'' ones (which is identical to the comoving observer ($t$ being the comoving time) modulo a redefinition of ``time'' $\eta=\int{dt}/{a(t)}$), are exposed to a radiation due to the gravitational particle creation, in the radiation dominated phase of the early universe. Although our analysis here is restricted to a two dimensional set up, there is a hope to extend it beyond two dimensions and in a fully four dimensional set up. We find the present study potentially interesting because, (i) it is technically simpler but has the same conceptual impact like the four dimensions, and (b)  it lays out a roadmap for a future four dimensional analysis. Note that such an (4D) analysis, if and when successfully carried out, might be compared with the CMB map just because the CMB might have taken place in the late part of the radiation dominated stage. %Therefore, this could well be the {\it only example} of particle creation phenomena that might be detected in background imprints of CMB radiation. 

In  \cite{modak1} we found new coordinates (we shall refer to them as $(T,~R)$ coordinates just for the sake of clarity of presentation) to describe the radiation dominated stage of the early universe by making a conformal transformation of the cosmological FRW coordinates. The use of $(T,~R)$ coordinates express the radiation dominated universe, following the inflationary stage, as a spherically symmetric, inhomogeneous spacetime which then offers a new, unitarily inequivalent field quantization for massless scalar fields. We also discussed several physical aspects related with the static and non-static observers in this spacetime. This discovery was followed by a systematic discussion of particle creation phenomena with respect to the {\it static observers} in this new spacetime who finds the so called {\it cosmological vacuum state} as containing particles. 

Here we concentrate on the other side, by selecting physical observers as our reference frame -- that is, we want to focus on the cosmological (and comoving) observers, whose frames are usually used for our understanding of cosmology. This case is much subtler in comparison to the above, simply because in the above case there exists a physical vacuum state (which is standard cosmological vacuum) and  on the contrary, here, we are not given a readily available physical vacuum state (at least in the radiation stage), that may appear as particle state, with respect to physical cosmological observers {\footnote{In the same sense in which the accelerated observers in the flat/Minkowski spacetime, due to their own motion, envision the Minkowski vacuum state as a particle excited state.}}. This is why we first define a new vacuum state, that comes into existence once we quantize the quantum field in the spherically symmetric form of background metric \cite{modak1} (we call it $T-$vacuum), prove its physical validity, and show that they are indeed particle excited state for physical cosmological observers.

Although, there is an analogy of our study of particle creation with the Unruh effect they, in fact, are not identical, and in a sense, quite complementary to each other. The main difference is that here, the two sets of coordinates, field modes and vacuum states are defined over the whole spacetime and not in a partial region of spacetime as in the Unruh case (e.g., Rindler vacuum). Second, in Unruh effect we identify ourselves as the Minkowski observer and therefore we don't have access to the radiation flux of the emitted particles which can only be accessed in the accelerated frame. However, in the present study, we (on the Earth) are neutral observers, neither in $T-$frame and nor in the comoving/cosmological frame and in fact non-inertial to both. This has a crucial advantage - any effect of particle creation, when observed either in comoving/cosmological frame or in $T-$frame won't just disappear when we move to our reference frame on the Earth! In fact, when we are able to calculate the particle creation due to $T-$vacuum for the comoving/cosmological observers, we can even translate the covariant quantities to the Earth's reference frame just by using coordinate transformations.  Notice that our non-inertial motion (in the Earth), with respect to the $T-$frame or comoving/cosmological frames, or the relative non-inertial motion between the comoving/cosmological frames, comes free just because of the expansion of the universe. This gives a slight hope for the observational verification (this question arises, only when we can extend this study for four dimensions). Lastly, what makes the second point viable is the well defined behavior of the $T-$ vacuum state (unlike the Rindler vacuum state) over the entire spacetime (which we also prove in this study for two dimensions). Our motivation of this study is to use the ``eyes'' of the comoving/cosmological observer and study physical effects due to the $T-$vacuum state which can be observed only in these frames. This is because we want to believe the standard folklore that, in Cosmology, comoving/cosmological observers are preferential. However, there is another possibility, that is using the ``eyes'' of $T-$observers and see the physical effect due to the cosmological vacuum state. A part of this calculation (only the excitation number density) was done before in \cite{modak1}, but we have a very little idea if there is any advantage of using the $T-$ observers' reference frame to understand cosmology and future studies are required to answer this question properly. But one thing, that is clear, is that, covariant quantities even in the $T-$frame, due to particle creation process, can be transformed to our frame on the Earth and there is no reason to believe that they will vanish just because of the coordinate transformation. Would that be of any interest? However, one must keep in mind that this study is being carried out for the radiation dominated stage, so any radiation/flux coming to us must travel throughout the subsequent expansion phases which may dilute the practical measure substantially. In any case, there are a few interesting outstanding aspects which we hope to deal with in future and this paper is a suitable beginning for such a dialogue.

The organization of the paper is as follows. In section \ref{s1} we give a brief review of the Unruh effect in two dimensions. In section \ref{s2} we give a short summary of mathematics that leads us to the $(T,~R)$ form of the radiation dominated universe. None of the above two sections are new but they are important to set the stage for the remaining study. The new results are reported in the remaining sections. Specifically,  in section \ref{s3} we discuss the observer dependent horizons and the special status enjoyed by the physical cosmological observers. We also build a platform to discuss quantum field theory in curved space by foliating the spacetime both with constant time-slices (both constant conformal time ($\eta$) and constant new time ($T$) time-slices) as well as with constant space-slices (both constant FRW space ($r$) and constant new space (R) space-slices).  Section \ref{s4} then discusses the particle creation phenomena and computes the particle number density. The next section \ref{s5} is used for calculating various components of the renormalized energy-momentum tensor (REMT), first in cosmological and then in the comoving frame, using the $T$-vacuum. We make a detailed comparison of this phenomena with the Unruh effect in section \ref{s6}. Finally in \ref{s7} we conclude.

%%%%%%%
\section{The Unruh effect} \label{s1}
Unruh effect is a well known example where one finds non-unique definition of quantum vacuum states. This effect is observed by an accelerated observer in Minkowski spacetime, who are a class of non-inertial observers, and who finds the Minkowski vacuum as a particle excited state. Inertial observers, on the other hand, related with each other by Lorentz transformations, have a unique definition of the vacuum state (the so called Minkowski vacuum). Note that, Lorentz transformations are essentially  linear between two sets of coordinates $(x,t)$ and $(x',t')$ and this plays a key role for the uniqueness of the vacuum state. On the other hand, one can also make a non-linear transformation, to the Minkowski metric 
\begin{equation}
ds^2 = dt^2 - dx^2,
\label{ms}
\end{equation}
of the form 
\begin{eqnarray}
t =  \frac{1}{\alpha} e^{\alpha \xi_1} \sinh(\alpha \xi_0)\label{ri1}\\
x = \frac{1}{\alpha} e^{\alpha \xi_1} \cosh(\alpha \xi_0) \label{ri2}
\end{eqnarray}
to express the spacetime as
\begin{equation}
ds^2 = e^{2\alpha\xi_1}\left(d\xi_0^2 - d\xi_1^2 \right)
\label{rs}
\end{equation}
which is called the Rindler spacetime and $(\xi_0, \xi_1)$ are the Rindler time and space coordinates. The original coordinates have limts $-\infty \le (t,x) \le \infty$. Rindler coordinates, $-\infty \le (\xi_0, \xi_1) \le \infty$, however, only cover the one-fourth of the Minkowski spacetime (the domain $x>|t|$). Lines of proper distance $\xi_1$ and constant Rindler time $\xi_0$ are plotted in $t-x$ plane.  
\begin{figure}[t]
\centering
%\rotatebox{270}{
\includegraphics[angle=0,width=4cm,keepaspectratio]{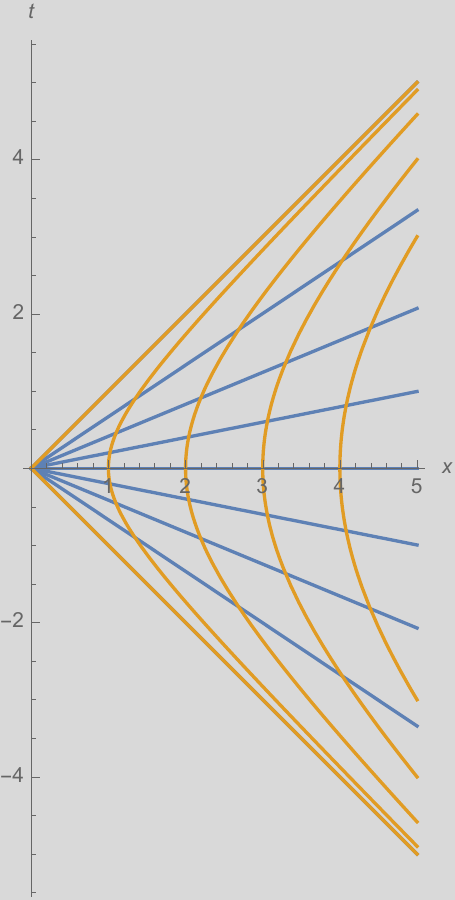}
\caption[]{(Color online) \begin{footnotesize}{Trajectories of constant Rindler timeslice $\xi_0 = \frac{1}{\alpha}\tanh^{-1}{(t/x)} = \text{const.}$, constant Rindler spaceslice $\xi_1 = x^2 - t^2 = \text{const.}$, along with the past and future Rindler horizons at $t=-x$ and $t=x$. Rindler coordinates only cover one-quarter of the Minkowski spacetime.}\end{footnotesize}}
\label{rindler}
\end{figure}

Construction of QFT in the spacetime \eqref{ms} and \eqref{rs} is simple. The Klein-Gordon equation for a massless scalar field takes the form
\begin{eqnarray}
\partial_u\partial_v\Phi = 0,\\
\partial_{\tilde{u}}\partial_{\tilde{v}}\Phi = 0,
\end{eqnarray}
in the null frame, $u=t-r,~ v=t+r$ and ${\tilde u}=\xi_0 - \xi_1~ {\tilde{v}}= \xi_0 + \xi_1$. Field solutions have left and right moving modes decoupled from each other and the field operator can be expressed as
\begin{eqnarray}
 \hat{\Phi} &=& \int_o^{\infty} \frac{d\omega}{\sqrt{4\pi\omega}}(e^{-i\omega u} a_\omega + e^{i\omega u} a_\omega^{\dagger} + \text{right moving}) \nonumber\\ \label{u1}\\
 &=& \int_o^{\infty} \frac{d\tilde{\omega}}{\sqrt{4\pi\tilde{\omega}}}(e^{-i \tilde{\omega} {\tilde u}} b_{\tilde{\omega}} + e^{i \tilde{\omega} {\tilde u}} b_{\tilde{\omega}}^{\dagger} + \text{right moving}) \nonumber\\ \label{u2}.
\end{eqnarray}
The definition of Minkowski and Rindler vacuum states are given by $a_{\omega}|0_M\rangle$ and $b_{\tilde{\omega}}|0_R\rangle = 0$. As is well known, both of these vacuums are well-defined in their respective frames -- if the quantum field is in $|0_M\rangle$ the particle detector remain unexcited for the static observer, and on the other hand, the detector remains unexcited for the accelerated observer only if the quantum field is in the state $|0_R\rangle$. It should be noted that $|0_R\rangle$ is only well defined where Rindler coordinates are well defined, i.e., one-quarter of the Minkowski spacetime ($|x| > t$) and it is singular on the Rindler horizons $u=0$ and $v=0$. This makes the Rindler vacuum unphysical for any observer which has access to the full spacetime - such as the Minkowski observer. The previous statement can be proved, \cite{muk} (pp. 106), where the relationship between the REMTs in two vacuums are related by 
\begin{equation}
\langle 0_{R}| T_{uu} |0_{R}\rangle = \frac{1}{\alpha^2u^2} \langle 0_{M}| T_{uu} |0_{M}\rangle,\label{relt}
\end{equation}
where $\alpha$ is the constant proper acceleration. Notice that the coordinate dependent prefactor on the right hand side of the above equation diverges at the future (Rindler) horizon at $u=0$. Since $|0_{M}\rangle$ is well defined,  it makes the Rindler vacuum a singular state for Minkowski observers who do not see any radiation coming from it. Similarly, for $T_{vv}$ one can prove that the left hand side diverges at the past Rindler horizon $v=0$.

There is no such issue of divergence for the Rindler observer looking into the Minkowski vacuum state, simply because $|0_{M}\rangle$  is  well defined in the Rindler wedge (as in other parts of the spacetime). One can also calculate the particle number density, for the Rindler observers, calculated in $|0_{M}\rangle$\footnote{The following result is true for a Minkowski space with finite volume, i.e., by ignoring a infrared divergence which appears from the infitite volume of the space \cite{muk}.}, 
\begin{equation}
\langle n_{\tilde\omega} \rangle= \frac{1}{e^{\frac{2\pi\tilde\omega}{\alpha}} - 1},
\end{equation}
which therefore tells us massless particles detected by the accelerated detector in Minkowski vacuum obey the Bose-Einstein distribution with the Unruh temperature $T=\frac{\alpha}{2\pi}$. Physically speaking, this particle excitation is not sourced by the vacuum fluctuations in the Minkowski spacetime itself, rather these fluctuations serve only as a mediator which can borrow the energy from the source responsible for the acceleration of the detector. However, {\em the energy of the detected particles are exponentially suppressed compared with the energy spent by the accelerating agent}\footnote{Mathematically this is related with the coordinate transformations relating the Minkowski and Rindler metrics, which are, for the null coordinates, exponentially related with each other.}. This make the Unruh effect quite unlikely to be detected directly.

The basic set up described above can be extended to four spacetime dimensions in a straightforward way - it does not need any new conceptual input. In the remaining part of the paper we will show that there is an effect of particle creation for physical observers, just because of their motion, in the radiation dominated early universe, and make an analogy with the Unruh effect.

%%%%%%%%%%

\section{New coordinates describing the radiation dominated universe}\label{s2}
Here, we provide a brief review of our earlier work \cite{modak1} related with  the installation of a new coordinate system in the radiation dominated universe. Let us start from the spatially flat FRW metric in comoving frame
\begin{equation}
ds^2= dt^2 - a^2(t)[dr^2 + r^2(d\theta^2 +{\sin^2{\theta}}~ d\phi^2)],
\label{frwcm}
\end{equation}
which in cosmological coordinates $(\eta,r, \theta, \phi)$, where $\eta = \int \frac{dt}{a(t)}$, is given by
\begin{eqnarray}
ds^2=a^2(\eta)[d\eta^2 - dr^2 - r^2(d\theta^2 +\sin^2{\theta}~ d\phi^2)].
\label{cffrw}
\end{eqnarray}
The scale factor $a(t)$ is an exponential function of time in the inflationary and dark energy dominated stages, whereas, it satisfies the power law equation $a(t) = a_0t^n$ for other stages of expansion, specifically, $n = 1/2$ for the radiation dominated stage and $n = 2/3$ for the matter dominated stage. The constant $a_0 = \sqrt{2{\cal H} e}$ (where ${\cal H}$ is the inflationary Hubble constant) for a universe starting from inflation and transiting into the radiation stage \cite{Singh:2013bsa}.  In the light-cone gauge $u = \eta-r$, $v=\eta+r$ and $r=\frac{v-u}{2}$ \eqref{cffrw} becomes
\begin{equation}
ds^2 = a^2 du dv - \frac{(v-u)^2}{4}a^2 (d\theta^2 + \sin^2\theta d\phi^2).\label{ncf}
\end{equation}
In \cite{modak1} we showed that, if we make a conformal transformation of the cosmological null coordinates, for a general functional dependence of $a(t)$, then it is only for the radiation dominated case where the resulting spacetime poses important symmetries that allow more than one (unitarily inequivalent) field decompositions. In fact, for $a(t) \propto t^{1/2}$, a conformal transformation of null coordinates of the following form \cite{modak1}
\begin{eqnarray}
U = T-R = \pm \frac{{\cal H}e}{2} u^{2} ~;~ V =T+R = \frac{{\cal H}e}{2} v^{2} 
\label{trans}
\end{eqnarray}
(where $+$ and $-$ sign applies for $u\ge0$ and $u\le 0$ respectively) takes the above metric into a spherically symmetric form. The full spacetime is now a direct sum of the following two spacetime metrics applicable to the sub-Hubble and super-Hubble regions \cite{modak1}
\begin{eqnarray}
d{s}^2 = &F_I(T,R) (dT^2 -dR^2) \nonumber \\ &- R^2 d\Omega^2,~ (\text{for}~ U \ge 0; T\ge R)
\label{mRT}
\end{eqnarray}
with
\begin{equation}
F_I(T,R) = \frac{(\sqrt{T+R} + \sqrt{T-R})^2}{4\sqrt{T^2 - R^2}}. \label{FTR1}
\end{equation}
\begin{eqnarray}
d{s}^2 = &F_{II}(T,R) (dT^2 -dR^2) \nonumber \\ &- T^2 d\Omega^2,~ (\text{for} ~U \le 0; ~T\le R)
\label{mRT1}
\end{eqnarray}
with
\begin{equation}
F_{II}(T,R) = \frac{(\sqrt{R+T} - \sqrt{R-T})^2}{4\sqrt{R^2 - T^2}}. \label{FTR2}
\end{equation}
We shall denote the sub-Hubble and super-Hubble regions (shortly we shall clarify this nomenclature), described by the metrics \eqref{mRT} and \eqref{mRT1}, as regions I and II, respectively.  In region I,  the new time and space coordinates are related with the cosmological time and space coordinates as
\begin{eqnarray}
T &=& (V+U)/2 = \frac{{\cal H}e}{2}(\eta^2 + r^2) \label{RT2}\\
R &=& (V-U)/2 = {\cal H}e \eta r \label{RT1}.
\end{eqnarray}
In region II, the relationships between these two sets of coordinates are reversed, so that
\begin{eqnarray}
T &=& (V+U)/2 = {\cal H}e \eta r \label{ter}  \\
R &=& (V-U)/2 =  \frac{ {\cal H}e}{2}(\eta^2 + r^2) \label{uf1}\label{rer}.
\end{eqnarray}
Notice that the above set of coordinate transformations are not linear (i.e., Lorentz type). These are non-linear transformations  (like Minkowski$\leftrightarrow$Rindler), although the functional form here are different than the Rindler case \eqref{ri1} and \eqref{ri2}.

By expressing the conformal factors $F_{I/II}(T,R)\rightarrow F_{I/II}(H,R)$, (where $H = (\frac{{\dot{a}}}{a})_{RD}$ is the Hubble parameter for radiation stage) we get \cite{modak1}  $F_I(H,R) = \frac{1}{1-H^2R^2}$ and $F_{II}(H,R) = \frac{1}{H^2T^2-1}$. Thus the light-cone boundary for the new observers at $T=R$ is nothing but the comoving Hubble radius at $R=1/H$. That justifies the above nomenclature for the metrics as sub and super-Hubble. 

It is important to note that $T$ remains timelike for both regions, I and II. There is no change of signature in \eqref{mRT} and \eqref{mRT1}. However, in the sub-Hubble region, the radius of the two sphere in \eqref{mRT}  (given by $R$) remains unchanged with time, for a static observer in $(T,~R)$ frame. The universe appears to be static, whose radial size does not change, for this observer. But there is a conformal change in the $R-T$ sector of the metric because of the conformal time dependent factor. The root of this radial staticity lies in the fact that the rate of expansion of the universe with respect to the cosmological time coincides with the motion of the observer which is static in the new spacetime (only in region I) but moving in cosmological frame. However, for the super-Hubble region (II) \eqref{mRT1}, a static observer (at constant $R$) in the new spacetime, does see the universe expanding with time $T$ since now the radius of the two-sphere is time-dependent  (in fact it is given by $T$). 

\begin{figure}[t]
\centering
\includegraphics[angle=0,width=10cm,keepaspectratio]{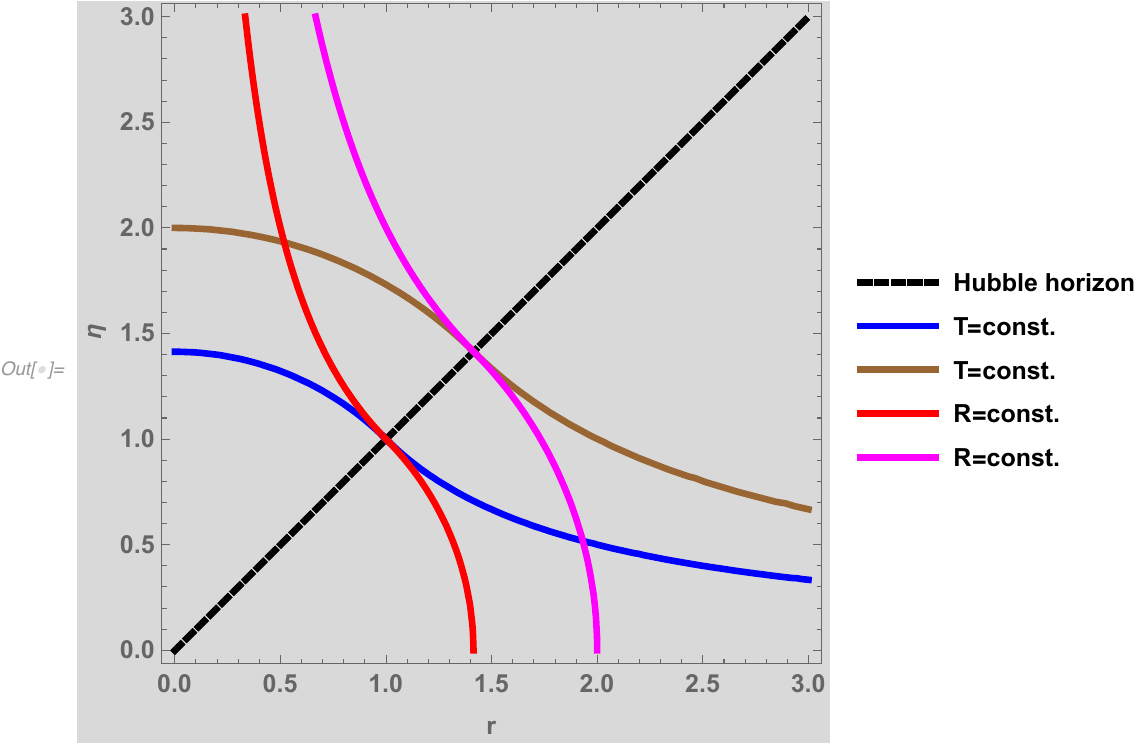}
\caption[]{(Color online) \begin{footnotesize}{Constant static time ($T$) and space ($R$) slices in the cosmological $(\eta,~r)$ plane (in the unit of ${\cal H} e =1$).  The $T=$ const. slices in the sub-Hubble region are circular due to \eqref{RT2} while they varies inversely with $r$ in the super-Hubble region due to \eqref{ter}. The $R=$ const. slices behaves exactly in a reverse manner due to \eqref{RT2} and \eqref{rer}. The intersection points are located on the light-cone boundary (at the Hubble scale) and there is no horizon for the static observer at the Hubble scale. The comoving Hubble radius is shown in black dotted line. Note that they provide a well-defined Cauchy slicing of the full spacetime.}\end{footnotesize}}
\label{fig-1}
\end{figure}

The constant $T$ and $R$ slices in $(\eta,~r)$ plane is plotted in Fig. \ref{fig-1}. Constant $R$ trajectories (i.e.,  the static observers in $(T,R)$ frame) are freely moving in cosmological frame (i.e., parallel to the $\eta$ axis) in the asymptotic past and future. However, they have acceleration and deceleration in between. With respect to the cosmological time,  they start accelerating in the super-Hubble region and reach luminal velocity to reach the Hubble radius in a finite timescale. Once they reach the Hubble radius they start decelerating, thus leaving the light trajectory to become sub-Hubble, and finally become indistinguishable from the freely moving observer in cosmological frame in the asymptotic future. Since this motion includes acceleration/deceleration there is a possibility of particle creation phenomena which was discussed in \cite{modak1}.  Now that we have a clear idea of the trajectory of a static observer in ($T,~R$) frame in the cosmological ($\eta,~r$) frame, we move to the next section where we study the inverse case, i.e., the trajectory of the cosmological observers, (who are static in cosmological ($\eta,~r$) frame), in the ($T,~R$) frame.

%%%%%%%%%%%%%
\section{Cosmological observers and a new spacetime foliation}\label{s3}

Let us start by considering the sub-Hubble metric \eqref{mRT} and a worldline given by $T=G(R)$. Along this worldline the ($R,T$) sector of the metric \eqref{mRT} becomes $ds^2 = \Xi  dR^2$, where the conformal factor
\begin{equation}
\Xi = \frac{(\sqrt{G(R)+R} + \sqrt{G(R)-R})^2}{4\sqrt{G(R)^2 - R^2}} \times (G'^2(R) -1).
\label{xi-1}
\end{equation}
If this factor diverges for some allowed value of $R>0$ and for a given $T=G(R)$ that will indicate the presence of horizon for the observer satisfying the aforementioned trajectory.  We are now free to chose any observer trajectory and test if there will be a horizon or not for the concerned observer. 

\subsection{Observer with constant radial velocity}
First, we consider a linear trajectory $T=G(R) = \alpha_0 R + \beta_0$ where $\alpha_0 \ge 1$ (necessary for region I) and $\beta_0 >0$ are dimensionless constants. This will mean that the radial velocity of the observer is constant $dR/dT ={1}/{\alpha_0}$ and thus they have no acceleration{\footnote{In general relativity, observers do not depend on the choice of the coordinates and there is an elaborate procedure for obtaining their trajectories using any coordinate system. Observer's four velocity, for a radial motion ($t,r$ frame), is $u^{\alpha}\partial_\alpha = \dot{t}\partial_t + \dot{r}\partial_r$, where dots are derivatives with respect to proper time $\tau$ of the observer. Using $T,R$ coordinates will only mean $u^{\alpha}\partial_\alpha = \dot{T}\partial_T + \dot{R}\partial_T$. It is then possible, to solve for its trajectory using a coordinate system and them transform it to any other coordinate systems using coordinate transformations. We are not doing this procedure here and directly writing such a relation for the sake of practical advantage in the present study.}}. In this case it is easy to check that $\Xi$ never diverges. {\it Therefore, any observer with constant radial velocity in ($T,~R$) frame do not encounter a horizon}. This is somewhat analogous to the result in Minkowski spacetime where inertial observers with constant velocity do not encounter horizon. 

\subsection{Observer with constant radial acceleration}
Next, let us consider a trajectory in region-I given by
\begin{eqnarray}
T=G(R) = \frac{1}{\alpha_0}(R+\beta_0)^{1/2},
\label{conac}
\end{eqnarray}
where $\alpha_0$ is a dimensionless constant and $\beta_0$ is a dimensionful constant. This trajectory represents an observer with constant radial acceleration $\frac{d^2R}{dT^2} = 2\alpha_0^2$. Using \eqref{xi-1} it is straightforward to see that this observer will encounter at most two event horizons at $R=\frac{1}{2\alpha_0^2} \pm \sqrt{\frac{1}{4\alpha_0^4} + \frac{\beta_0}{\alpha_0^2}}$ {\footnote{There will be only one horizon with `+' sign before the square-root if $\beta_0$ is positive, whereas, for the contrary with negative $\beta_0$ there will be two horizons both for $R>0$.}}. The appearance of horizon for an accelerated motion is reminiscent of the case for accelerated observers (Rindler observers) in the Minkowski spacetime who finds event (Rindler) horizon in the Minkowski spacetime and are confined to one-fourth of the full spacetime. The crucial difference of that with the present situation being, while the accelerated observer in Minkowski spacetime can be easily imagined and attributed to a class of physical observers, here the observers with a trajectory \eqref{conac} has a little physical motivation -- it is not clear if they are of physical interest to us in the understanding cosmology.

This situation, however, will change dramatically in the next sub-section where we will attribute a trajectory to a class of physical observers in cosmology (observers with proper time registered as the conformal time $\eta=\int{dt}{a(t)}$).

\subsection{Cosmological Observers}
Let us consider the worldline in the sub-Hubble region-I
\begin{equation}
T=G(R) = \alpha_1 R^2 + \beta_1,
\label{cs2}
\end{equation} 
where $\alpha_1$ and $\beta_1$ are dimensionful positive definite constants. These observers have a position dependent radial velocity $\frac{dR}{dT} = \frac{1}{2\alpha_1 R}$ and a nonzero, position-dependent radial deceleration 
\begin{equation}
\frac{d^2R}{dT^2} = - \frac{1}{4\alpha_1^2 R^3},
\label{radac}
\end{equation}
whose magnitude is increasing as the observer approaches the center $R=0$. That is, they tend to come to a rest close to the centre and become freely moving observer at their causal future. As evident from \eqref{cs2} these observers follow a parabolic trajectory in $(T,R)$ plane. Substituting this in \eqref{xi-1} we find
\begin{eqnarray}
\label{xi}
&&\Xi \notag \\
&&= \frac{(2\alpha_1 R+1)(\sqrt{(\alpha_1 R +1) R + \beta_1} + \sqrt{(\alpha_1 R +1) R - \beta_1})^2}{4\sqrt{\alpha_1 R^2 + \beta_1 + R}}   \nonumber \\
&&\times \frac{2\alpha_1 R -1}{\sqrt{\alpha_1 R^2 + \beta_1 - R}}.
\end{eqnarray}
The first factor is always finite, so it is only the second factor which determines if $\Xi$ diverges. Clearly there are two divergences for the root $R_0 = \frac{1\pm\sqrt{1-4\alpha_1\beta_1}}{2\alpha_1}$ if $\beta_1<1/4\alpha_1$ implying, therefore, existence of two horizons. If $\beta_1$ vanishes there is only one horizon at $R_0 = 1/\alpha_1$. Further, there is a very special case for  $\beta_1 = 1/4\alpha_1$ for which the second factor in \eqref{xi} is $\sqrt{4\alpha_1}$. Therefore, among all observers satisfying \eqref{cs2}, it is this and only this observer satisfying $4\alpha_1\beta_1=1$ does not encounter any horizon anywhere in the spacetime. 

It is genuinely interesting to show that the above special observers, satisfying \eqref{cs2} (with $\beta_1 = 1/4\alpha_1$),  are none other than the cosmological observers at any position $r$ in region-I. To demonstrate that we use the relations \eqref{RT2} and \eqref{RT1}. We find that the constant $r=r_0$ trajectories in $R-T$ plane satisfy an identical relationship like $T= \alpha_1 R^2 + \beta_1$ with $\beta_1 = 1/4 \alpha_1 =  {\cal H} e \eta_0^2/2$. In fact a constant $r$ trajectory also satisfies the same relationship just because \eqref{RT2} and \eqref{RT1} are symmetric under the interchange of $r$ and $\eta$. Therefore, {\it cosmological observers enjoy a very special status in ($T,~R$) frame --- they are the only observers who are accelerating radially satisfying \eqref{radac}, but do not encounter any horizon (due to a coordinate singularity)}. It, therefore, needs no more justification to say that neither the comoving observers with proper time $t$ will encounter a horizon in the whole spacetime. This is of course expected for the cosmological observers as well as for the comoving observers who are after all free to communicate with all spacetime events respecting causality. 

Now let us turn our attention to the super-Hubble region-II. Everything that we have showed above for the worldline \eqref{cs2} also apply to the following worldline in region-II
\begin{equation} 
R=G(T) = \alpha_1 T^2 + \beta_1,
\label{cs2-sh}
\end{equation}
These observers correspond  to the trajectory of the cosmological observers at constant $r=r_0$ with the identification $\beta_1 = 1/4 \alpha_1 =  {\cal H} e r_0^2/2$ in the $T,~R$ frame. Also, notice that at $T=R$ \eqref{cs2} and \eqref{cs2-sh} are identical. These observers have a time dependent velocity $\frac{dR}{dT} = 2\alpha_1 T$ and a constant acceleration $\frac{d^2R}{dT^2} = 2\alpha_1$ in the region-II.

Now looking at the whole picture, we see that at the asymptotic past (i.e., the beginning of the radiation dominated universe $\eta = \eta_{rd}$) a cosmological observer at $r=r_0$ is freely moving in region-II and subsequently it has the following interesting trajectory in the ($T,~R$) frame: a cosmological observer starts accelerating at a constant rate $2\alpha_1={\cal H} e r_0^2/2$ as the time $\eta$ increases and during this period the radial velocity also increases at a rate $2\alpha_1 T$. This is continued until the cosmological time $\eta = r_0$ or at the light-cone boundary. At this point the radial velocity $\frac{dR}{dT} = 2\alpha_1 T = 1$, i.e., it reaches the luminal velocity with respect to a freely moving observer in $(T,~R)$ frame. After this the cosmological observer enters in region-I  where it starts decelerating with respect to a static $R$ observer satisfying \eqref{radac} and coincides with the freely moving observer in $(T,~R)$ frame.

%One can, in fact, go on to discuss other observer trajectories but we rather want to turn our entire focus on the FCOs.

%%%%%%%%%%%%%%%%%%

\subsection{New spacetime foliation considering cosmological observers}

Since we are about to construct a consistent quantum field theory keeping cosmological observers in focus, it is customary to show that there exists such a construction, and the first step in doing that is to foliate the spacetime in terms of the spacelike and timelike hypersurfaces.  Particularly, we want to see how the constant $\eta$ (which defines a ``time-slice'' or a spacelike hypersurface) and constant $r$ (which defines a ``space-slice'' or timelike hypersurface) surfaces will look like. Since the relationships between the two sets of coordinates, in \eqref{RT2} and \eqref{RT1}, are symmetric under the exchange of $\eta$ and $r$ (so is true for \eqref{ter} and \eqref{rer}) we have to be rather cautious to identify the time-slices and the space-slices. Below we provide a valid construction.

\begin{figure}[t]
\centering
%\rotatebox{270}{
\includegraphics[angle=0,width=8cm,keepaspectratio]{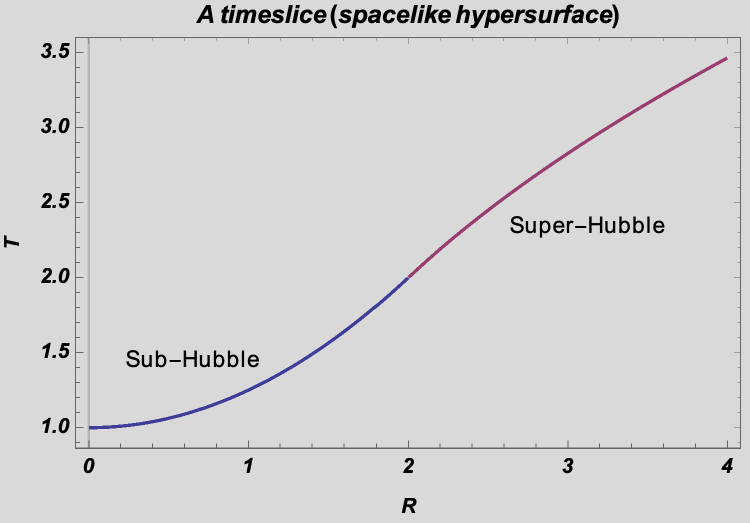}
\caption[]{(Color online) \begin{footnotesize}{A constant cosmic time ($\eta= \eta_0$) slice in the $(T,~R)$ plane (the depicted slice has $\eta_0 = \sqrt{2/{\cal H}e}$).  In the sub-Hubble region it consists of the lower portion (with respect to the SLR) of the parabola \eqref{pab1} (showed by the blue curve), while in the super-Hubble region is foliated by the parabola \eqref{pab2}  (only the portion right to the SLR --- showed by the purple curve). They intersect at the Hubble radius at $T=R$ where \eqref{pab1} and \eqref{pab2} are identical. This is an example of a well-defined spacelike hypersurface (each point on this line is a two sphere) where one can assign a Cauchy data.} \end{footnotesize}}
\label{foli-1}
\end{figure}

First, let us plot the $\eta = \eta_0$ timeslices. This follow from the relationships \eqref{RT2}   and \eqref{RT1}. Substituting $r$ from \eqref{RT1} into \eqref{RT2} expresses $T=T(R,\eta)$ and then a constant $\eta$ slice will be the parabola 
\begin{equation}
T = \frac{R^2}{2 {\cal H} e \eta_0^2} + \frac{1}{2}{\cal H} e \eta_0^2 
\label{pab1}
\end{equation}
defined for the sub-Hubble region-I. We only consider the portion of this parabola that is below the  Semi-Latus-Rectum (SLR). The SLR, the parabolic curve \eqref{pab1} and the Hubble horizon meet at one point. From that point onwards we extend the timeslice to the super-Hubble region-II. In order to do that now we consult the relationships \eqref{ter} and \eqref{rer}. We substitute $r$ from \eqref{ter} and express $R=R(T,\eta)$ and then the timeslice will simply be a portion of the parabola
\begin{equation}
R = \frac{T^2}{2 {\cal H} e \eta_0^2} + \frac{1}{2}{\cal H} e \eta_0^2
\label{pab2}
\end{equation}
for the super-Hubble region-II. This portion is chosen as the the parabolic curve on the right of SLR of \eqref{pab2}. The complete $\eta=const.$ hypersurface is now given by the union of the lower portion (of the SLR) of the parabola \eqref{pab1} and the right portion (of the SLR) of the parabola \eqref{pab2}. This arrangement is plotted mathematically in Fig. \ref{foli-1}.

\begin{figure}[t]
\centering
%\rotatebox{270}{
\includegraphics[angle=0,width=8.5cm,keepaspectratio]{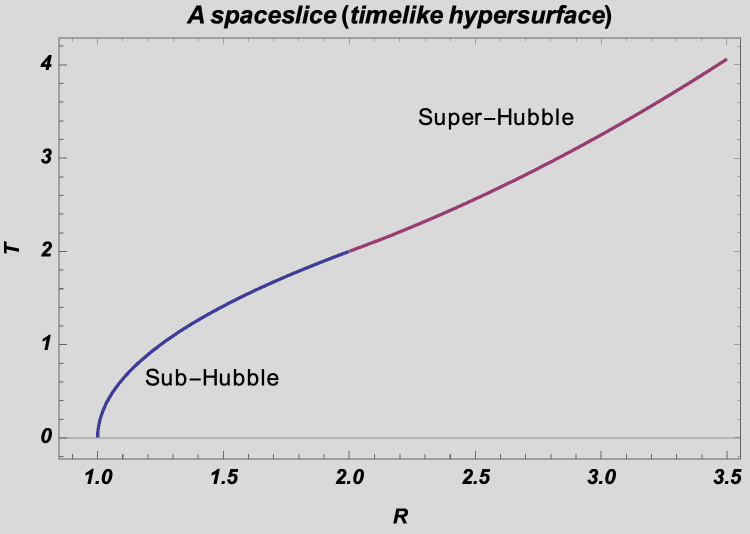}
\caption[]{(Color online) \begin{footnotesize}{A constant cosmic space ($r=r_0$) slice in the cosmological $(T,~R)$ plane (the depicted slice has $r_0= \sqrt{2/{\cal H} e}$).  The $r=r_0$ slice in the sub-Hubble region is the upper portion to the SLR of the parabola \eqref{pab3} (showed by the blue curve), while the super-Hubble region is foliated by the parabola \eqref{pab4}  (only the portion left to the SLR, showed by the purple curve). They also intersect at the Hubble radius at $T=R$ where \eqref{pab3} and \eqref{pab4} are identical parabolas. This is the worldine of the cosmological static observer at $r=r_0=\sqrt{2/{\cal H} e}$ while looked upon in $T,~R$ frame.}\end{footnotesize}}
\label{foli-2}
\end{figure}

In a similar way we can construct the timelike hypersurfaces or ``spaceslices'' with $r=const.=r_0$. For that, in region-I we now substitute  $\eta$ from \eqref{RT1} into \eqref{RT2} expresses $T=T(R,r)$
\begin{eqnarray}
T = \frac{R^2}{2 {\cal H} e r_0^2} + \frac{1}{2}{\cal H} e r_0^2, 
\label{pab3}
\end{eqnarray}
and in region-II we substitute $\eta$ from \eqref{ter} and express $R=R(T,r)$, given by
\begin{eqnarray}
R = \frac{T^2}{2 {\cal H} e r_0^2} + \frac{1}{2}{\cal H} e r_0^2.
\label{pab4}
\end{eqnarray}
Notice that they are identical to the other set \eqref{pab1} and \eqref{pab2} under the exchange of $r$ with $\eta$. Now the $r=const.$ spaceslices are constructed by joining the upper portion (of the SLR) of the parabola \eqref{pab3} with the left portion (of the SLR) of the parabola \eqref{pab4} which will cover the whole regions of spacetime (I and II) covering the sub and super-Hubble scales. These slices are shown in Fig. \ref{foli-2}.

\begin{figure}[t]
\centering
%\rotatebox{270}{
\includegraphics[angle=0,width=8.5cm,keepaspectratio]{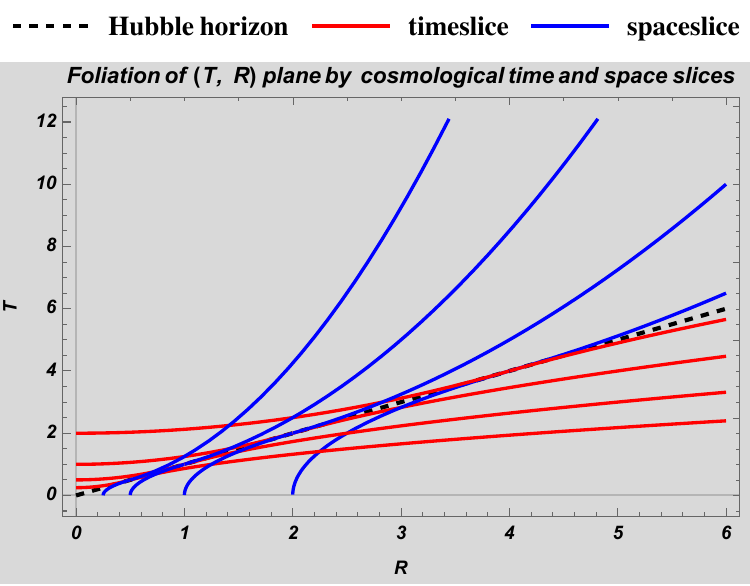}
\caption[]{(Color online) \begin{footnotesize}{A complete foliation by constant time and space slices in cosmological frame depicted in the $(T,R)$ plane. The black dotted line is the comoving Hubble radius. There is no horizon for cosmological observers at the Hubble scale or anywhere in the spacetime. This figure is analogous to Fig. \ref{rindler} but with a major difference -- here the space and time slices corresponding to the cosmological observer cover the full spacetime (as they should), while the Rindler space and time slices cover only one-quarter of Minkowski spacetime  (Fig. \ref{rindler}).}\end{footnotesize}}
\label{fig1}
\end{figure}

A combined plot including the spacelike and timelike hypersurfaces are shown in Fig. \ref{fig1}. These slices foliate the complete spacetime, (basically the $(T,R)$ plane; each point in this plane is a two sphere) with constant $\eta$ (timeslices) and constant $r$ (spaceslices), in a consistent manner so that the Cauchy problem is well posed. Recall that this foliation is independent of the other foliation discussed in Fig. \ref{fig-1}.

%provide tw o different foliation of the full spacetime by means of spacelike and timelike hypersurfaces. 

%Unlike, Rindler and Minkowski case, here cosmological and Modak coordinates cover the complete patch of the spacetime and hence when we make field quantization the mode functions will be complete for both cases. This does not happen in Rindler-Minkowski case since the Rindler observers only cover one fourth of the spacetime due to coordinate singularity and therefore the mode functions in Rindler frame remain incomplete. But here, neither FCOs nor static observers in Modak frame encounter a horizon. This will have an important impact on the particle creation process. 

Notice also, that, once again, in Fig. \ref{fig1} the cosmological observers at $r=$const. (vertical curves) are, in fact, freely moving in $(T,~R)$ frame only in the asymptotic past and future, but they are accelerated (and decelerated) radially  in super (and sub) Hubble regions, respectively. cosmological observers attain the luminal velocity at the Hubble scale and this is exactly analogous to the case discussed in the last section for static observers in cosmological frame \cite{modak1} who attain the luminal velocity in $(T,~R)$ frame. In fact, we expect this pattern to be reciprocal because, they are, after all,  accelerated or decelerated with respect to each other. Another important thing to note is the fact that  an initial Cauchy data, on any initial space-like hypersurface, can be defined either on a $T=$const surface in Fig. \ref{fig-1} or on a $\eta=$const surface in Fig.\ref{fig1}. For both situations, there is a corresponding time translation operator, defined with respect to cosmological time $\eta$ or the new time $T$. In quantum theory corresponding Hamiltonian operators would be $\hat{H}_1 := i\hbar\frac{\partial}{\partial \eta}$ and $\hat{H}_2 :=i\hbar\frac{\partial}{\partial T}$ which will generate the time translations of the form $|\psi (\eta) \rangle := e^{-i \hat{H}_1 (\eta - \eta_0)} |\psi (\eta_0) \rangle$ and $|\psi (T) \rangle := e^{-i \hat{H}_2 (T - T_0)} |\psi (T_0) \rangle$.

%%%%%%%%%%%%%%%%%
\section{Particle creation}\label{s4}

Now we proceed on to another important part of our discussion which is gravitational particle creation for cosmological observers. In our earlier work \cite{modak1} we provided a detailed discussion of particle creation phenomena with respect to the static observer in $(T,R)$ spacetime who finds the cosmological vacuum as containing particles. As showed in Fig. \ref{fig-1}, these observers have non-trivial trajectory in FRW coordinates. Here we want to calculate the particle content for cosmological observers who are at $r=$const. and following the trajectories showed in Fig. \ref{fig1}. These observers will be exposed to a radiation due to their motion. This will be a consequence of a newly found vacuum state that we are about to introduce in this section. We shall restrict ourselves to a two dimensional set up which will keep our analysis simpler, yet, physically very intuitive. The four dimensional calculation using spherical polar coordinates is a bit more involved than a two dimensional analysis and for this case it needs to be handled with more caution, and will be reported in a future work \cite{modak2}.

\subsection{Number density in two dimensions}

In two dimensions (ignoring $\theta,~\phi$ coordinates) the field equations for the massless scalar field read $\partial_u\partial_v \Phi= 0$ for \eqref{cffrw} and $\partial_U\partial_V \Phi=0$ for both \eqref{mRT} and \eqref{mRT1}. The field operator, expanded in two bases as
\begin{eqnarray} 
 \hat{\Phi} &=& \int_o^{\infty} \frac{d\omega}{\sqrt{4\pi\omega}}(e^{-i\omega u} a_\omega + e^{i\omega u} a_\omega^{\dagger} + \text{right moving}) \nonumber\\ \label{exp1}\\
 &=& \int_o^{\infty} \frac{d\omega}{\sqrt{4\pi\Omega}}(e^{-i\Omega U} b_\Omega + e^{i\Omega U} b_\Omega^{\dagger} + \text{right moving}) \nonumber\\ \label{exp2}.
\end{eqnarray}
The Bogolyubov coefficients relating the annihilation operator $$a_\omega = \int_0^\infty d\Omega (\alpha_{\omega\Omega} b_\Omega - \beta_{\omega\Omega} b_\Omega^\dagger)$$ in terms of the sum of creation and annihilation operators of the other basis can be easily calculated as 
\begin{eqnarray}
\alpha_{\omega\Omega} &=& \frac{1}{2\pi}\sqrt{\frac{\omega}{\Omega}} \int_{-\infty}^{\infty} du e^{-i\Omega U +i\omega u}, \\
\beta_{\omega\Omega} &=& -\frac{1}{2\pi}\sqrt{\frac{\omega}{\Omega}} \int_{-\infty}^{\infty} du e^{i\Omega U +i\omega u}.
\label{beta}
\end{eqnarray}
The average particle number density for a given frequency is then given by
\begin{equation}
\langle n_{\omega} \rangle = \int_0^{\infty} d\Omega |\beta_{\omega\Omega}|^2
\label{n}
\end{equation}
where, $n_{\omega} = a_\omega^\dagger a_\omega$ is the number operator defined in the cosmological basis and the expectation value $\langle 0_{T} | n_{\omega} | 0_{T}\rangle$ is calculated in the vacuum state in the new basis, as defined  by $b_\Omega |0_{T}\rangle = 0$. We refer to this vacuum state  $|0_{T}\rangle$ as $T$-vacuum.

\begin{figure}[t]
\centering
%\rotatebox{270}{
\includegraphics[angle=0,width=8.5cm,keepaspectratio]{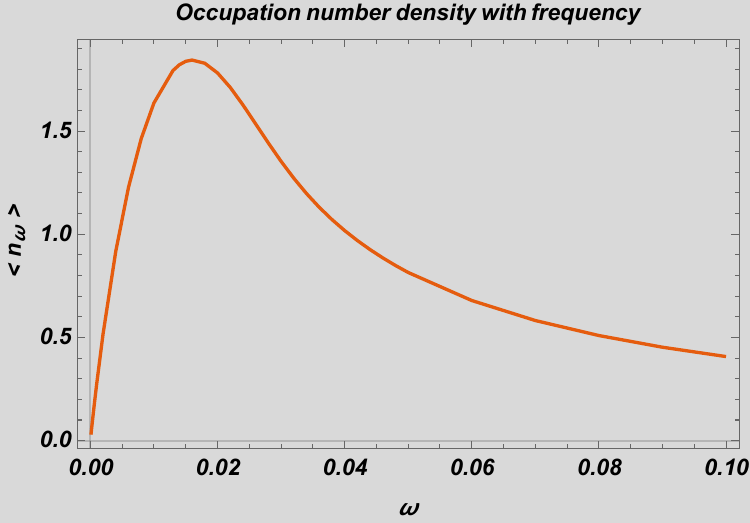}
\caption[]{(Color online) Plot of particle number density versus frequency for a two dimensional set up. This figure corresponds to ${\cal H} =1$ and introduced a infra-red cut-off at $\Omega = 0.001$.}
\label{fig2}
\end{figure}

To calculate the coefficient \eqref{beta} we first divide the integral for $u \le 0$ and $u\ge 0$ and use appropriate relationships relating two null coordinates as appear in \eqref{trans}. After performing the integration \eqref{beta} we can derive
\begin{equation}
|\beta_{\omega\Omega}|^2 = \frac{\omega }{8 e {\cal H} \pi^2 \Omega^2 } \left(1+ \sin(\frac{\omega^2}{\Omega e {\cal H}})\right)\Gamma^2[\frac{1}{2}, \frac{\omega^2} {2 \Omega e {\cal H}}] \label{beta1}
\end{equation}
where $\Gamma$ is an upper incomplete gamma function. Equation \eqref{n} then provides average particle number density. Unfortunately, it is difficult to get an exact analytical result for the particle number density \eqref{n} using \eqref{beta1}. We therefore use numerics and plot the number density in Fig \ref{fig2}. We have noted an infrared divergence in $\langle n_\omega \rangle$ which is unphysical and appears in other situations such as particle creation by the moving mirror \cite{BD}. This divergence is usually avoided just by making an infrared cutoff in $\Omega$ which is also the case in our case (Fig \ref{fig2}). This is a new result and quite significant from the physical perspective since it establishes the fact that the cosmological observers are exposed to a radiation just because of the existence of the new vacuum state $|0_{T}\rangle$ which can only be defined once we know the new coordinates \eqref{mRT} and \eqref{mRT1} and field theory based on these coordinates.

%%%%%%%%%%%%%
\section{ T-vacuum and renormalized energy-momentum tensor}
\label{s5}
In this section we discuss more about the new vacuum state that we found in the last section while quantizing the field equation in the new reference frame and use it to calculate various components of renormalized energy-momentum tensor applicable to the two dimensional set up discussed here.

%\subsection{The T-vacuum}
We have already seen equations \eqref{exp1} and \eqref{exp2} give rise two choices of vacuum states --- the standard cosmological vacuum state defined by $a_\omega |0_{C}\rangle = 0$ as well as a newly defined vacuum state which we call here the $T$-vacuum defined as $b_\Omega |0_{T}\rangle = 0$. An analogous situation arises for the Rindler-Minkowski case where one has the standard Minkowski vacuum and also the Rindler vacuum. As we have already shown in section \ref{s1}, that the Rindler vacuum state is unphysical since it needs an infinite energy to be produced. On the contrary, Rindler observers are physical observers and they do find the Minkowski vacuum as a thermal state and register particle excitations which is the main point of the Unruh effect \cite{unruh}.  

With a clear idea of the above, now we want to examine here if the newly defined $T$-vacuum is well-defined everywhere or it has similar problems like the Rindler vacuum. To do this we first take the expectation values of the operators $(\partial_u\hat{\Phi})^2$ and $(\partial_U\hat{\Phi})^2$ in the cosmological and $T$-vacuum states respectively. This can be calculated from the field expansions \eqref{exp1} and \eqref{exp2} which gives (simply because $|0_{C}\rangle$ is annihilated by $a_\omega$ and $|0_{T}\rangle$ by $b_\Omega$.)
\begin{equation}
\langle 0_{C}| (\partial_u\hat{\Phi})^2 |0_{C}\rangle = \langle 0_{T}| (\partial_U\hat{\Phi})^2 |0_{T}\rangle.
\label{emt}
\end{equation}
We can use the above equation and the transformation law for derivative operators to express the quantity 
\begin{eqnarray}
 \langle 0_{T}| \frac{1}{2}(\partial_u\hat{\Phi})^2 |0_{T}\rangle &=& \left(\frac{\partial{U}}{\partial{u}}\right)^2 \langle 0_{T}| \frac{1}{2}(\partial_U\hat{\Phi})^2 |0_{T}\rangle,\label{emt20}\\
 && = ({\cal H} e u)^2  \langle 0_{C}|  \frac{1}{2}(\partial_u\hat{\Phi})^2 |0_{C}\rangle, \label{emt2}
\end{eqnarray}
where for the second line we have used the relationship between $U(u)$ in \eqref{trans}, as well as equation \eqref{emt}. The above equation simply gives the relationship between the $uu$ component of the energy-momentum tensor ($\langle T_{uu} \rangle$) evaluated in the $T$-vacuum with that evaluated in cosmological vacuum in the following way
\begin{equation}
\langle 0_{T}| T_{uu} |0_{T}\rangle = ({\cal H} e u)^2 \langle 0_{C}| T_{uu} |0_{C}\rangle.
\label{reltc}
\end{equation}
An analogous relationship between the Rindler and Minkowski coordinates also holds for two dimensions \eqref{relt} which, as we already mentioned, proved the unphysical nature of the Rindler vacuum for Minkowski observers. In the same spirit it is necessary for us to consider \eqref{reltc} and check its legitimacy in the frame of a cosmological observer. If $|O_T\rangle$ shows singularities in the components of REMT in the cosmological observers frame we cannot accept the validity of the said vacuum state and there won't be any physical meaning of the particle creation process. Below we show that it is indeed a physical vacuum state even in the frame of a cosmological observer.

In order to calculate the l.h.s of \eqref{emt2}  we first need to know the REMT in the cosmological vacuum. Since we are limited to a two dimensional set up, we recall various studies by Birrel, Davies, Fulling, Unruh and Bunch in the middle to late seventies \cite{2d} which provide a covariant, divergence free  expression for REMT in the cosmological vacuum. This is given by
\begin{equation}
\langle T_{\mu\nu} \rangle_{C} = \langle 0_{C}| T_{\mu\nu} |0_{C}\rangle = \Theta_{\mu\nu} + \frac{R}{48\pi} g_{\mu\nu}
\label{2dem}
\end{equation}
where, $\Theta_{uv}=0$ and
\begin{eqnarray}
\Theta_{uu} = \frac{1}{24\pi} (\partial_u D_1 - \frac{1}{2}D_1^2)\\
\Theta_{vv} = \frac{1}{24\pi} (\partial_v D_2 - \frac{1}{2}D_2^2)\\
D_1 = \frac{1}{a^2} \partial_u (a^2),~ \& ~ D_2 = \frac{1}{a^2} \partial_v (a^2)
\end{eqnarray}
where $a (u,v) = \frac{{\cal H} e}{2}(u+v)$ is the scale factor. In our case, for the radiation dominated universe we have $R=0$, i.e., the second term in \eqref{2dem} is inevitably zero. That gives, $\langle T_{uv} \rangle_{C} = 0$. As for the others we get, using all of the above set of equations
\begin{equation}
\langle T_{uu} \rangle_{C} = \langle T_{vv} \rangle_{C} =  -\frac{1}{6\pi (u+v)^2}.
\label{tuu}
\end{equation}
Now substituting this in \eqref{reltc} we get
\begin{equation}
\langle 0_{T}| T_{uu} |0_{T}\rangle =\langle T_{uu} \rangle_{T}  = -\frac{{\cal H}^2 e^2 }{6\pi (1+v/u)^2}.
\label{reltc1}
\end{equation}
Similarly, one can easily calculate 
\begin{eqnarray}
\langle 0_{T}| T_{vv} |0_{T}\rangle =\langle T_{vv} \rangle_{T}  = -\frac{{\cal H}^2 e^2 }{6\pi (1+u/v)^2},
\label{reltc2}
\end{eqnarray}
whereas, the other components $\langle T_{uv} \rangle_{T} =0 = \langle T_{vu} \rangle_{T}$. These are extremely important results which prove the well-defined-ness of the $T$-vacuum. Notice that, none of the above expressions have singularity at the Hubble radius ($u=0$ or $T=R$) where two spacetime patches are glued together. Precisely, we have $\lim_{u\rightarrow 0} \langle T_{uu} \rangle_{T} = 0$ and $\lim_{u\rightarrow 0} \langle T_{vv} \rangle_{T} = - \frac{{\cal H}^2 e^2 }{6\pi }$. Also, for asymptotic values of $u \rightarrow \pm \infty$, we have $\lim_{u\rightarrow \pm \infty} \langle T_{uu} \rangle_{T} = - \frac{{\cal H}^2 e^2 }{6\pi }$, and $\lim_{u\rightarrow \pm \infty} \langle T_{vv} \rangle_{T} = 0$. On the other hand if we take similar limits on the advanced null coordinate, we get $\lim_{v\rightarrow 0} \langle T_{uu} \rangle_{T} = - \frac{{\cal H}^2 e^2 }{6\pi }$, $\lim_{v\rightarrow 0} \langle T_{vv} \rangle_{T} =0$, $\lim_{v\rightarrow \pm \infty} \langle T_{uu} \rangle_{T} =0$, and $\lim_{v\rightarrow \pm \infty} \langle T_{vv} \rangle_{T} = - \frac{{\cal H}^2 e^2 }{6\pi }$. The above analysis proves that the $T$-vacuum state that we introduce in this paper is well-defined and has a physical importance. Using \eqref{reltc1} and \eqref{reltc2} we can also calculate
\begin{eqnarray}
\langle 0_{T}| T_{\eta\eta} |0_{T}\rangle &=& T_{rr} = - \frac{{\cal H}^2 e^2}{12\pi} \left( 1+ \frac{r^2}{\eta^2} \right), \\
\langle 0_{T}| T_{\eta r} |0_{T}\rangle &=& T_{r \eta} = - \frac{{\cal H}^2 e^2}{12\pi} \left(\frac{r}{\eta}\right).
\end{eqnarray}

One of the important aspects of calculating covariant, renormalized energy-momentum tensor is that the result can be transformed in other coordinates which are suitable to the observers of interest. In cosmology the most important observer is identified as the ``comoving observer''.  Comoving observers are the nearest analogy to the inertial observers in flat space which are not acted upon by any external force. The only difference in the comoving observers and cosmological observers is the time coordinate --- FRW metric in the comoving frame is given by \eqref{frwcm}, whereas it is in the cosmological frame is given by \eqref{cffrw}. Transforming the above results in comoving frame we get
\begin{eqnarray}
\langle 0_{T}| T_{tt} |0_{T}\rangle &=& - \frac{{\cal H}^3e^3 t}{6 \pi} \left( 1+ \frac{{\cal H} e r^2}{2t} \right), \\
\langle 0_{T}| T_{rr} |0_{T}\rangle &=& - \frac{{\cal H}^2 e^2}{12\pi} \left( 1+ \frac{{\cal H}e r^2}{2t} \right) \\
\langle 0_{T}| T_{t r} |0_{T}\rangle &=& T_{r \eta} = - \frac{{\cal H}^2 e^2}{24\pi} \left(\frac{r}{t}\right).
\end{eqnarray}
The $tt$-component is the energy density that is turning out to be negative here, thus giving an effect of repulsive gravity, and in a full four dimensional set up this would mean an accelerated expansion (the radiation pressure, treated classically, is already driving the background spacetime) just because of this term. This is however a two dimensional toy model and generalising this to a four dimensional set up is yet to be done. Likewise, the other diagonal term ($rr$-component) is interpreted as pressure. The off-diagonal components represent momentum-densities which are equivalent to the components of linear momentum. Since it is two dimensional model there is no shear stress acting here. One can crosscheck the correctness of the above expressions by calculating the left hand side of the two-dimensional trace anomaly equation $g^{ab}\langle 0_{T} | T_{ab}|0_{T}\rangle = \frac{R}{24\pi}$. Indeed, the left hand side gives us zero which is compatible with the fact that  the Ricci scalar vanishes for the RD universe.

%We should note again that the above analysis is restricted so far for two dimensions and the interpretations can't be trivially generalized to a four dimensional set up.  We are hoping shortly extending this for a full four dimensional set up in a forthcoming paper \cite{modak2}. Nevertheless, this study should establish the physical importance of the new coordinates that we define here and more  so for the $T$-vacuum state.

%%%%%%%%%%
\section{Comparison with the Rindler-Minkowski case and Unruh effect}
\label{s6}

Even in the absence of gravity, the relationship between the Minkowski and Rindler metrics representing the flat spacetime has been a cornerstone for our understanding of QFT in curved space. Often discussions about particle creation in other cases, such as in black hole spacetimes, are compared with the Unruh effect. We intend to do the same with our discovery of the particle creation reported in the last section. This will help us to realize the importance of this study.

The first step is to compare the relationships between the coordinates. A comparison for the two dimensional case (angular coordinates are unchanged) is presented in the following tabular form:
 
\begin{center}
\begin{tabular}{|c| l | l | } \hline
 Item  & Rindler -- Minkowski & FRW -- New coord. \\ \hline
 & \begin{footnotesize}\end{footnotesize} & \begin{footnotesize}\end{footnotesize} \\
 
{ i) Metrics} & i) Rindler:  & i) FRW coord.:  \\
& $ds^2 = e^{2 \alpha \xi_1} [(d\xi_0)^2 - (d\xi_1)^2]$ & $ds^2 = a^2 (d\eta^2 - dr^2)$\\

 & Minkowski:    &  New coord.:  \\
 
 & $ds^2= dt^2 - dx^2$ & $d{s}^2 = F_{I/II}(T,R) (dT^2 -dR^2)$ \\
 &&\\
{ ii) Relationships} & ii) $t = \frac{e^{\alpha\xi_1}}{\alpha} \sinh(\alpha\xi_0)$ & ii) $T =  \frac{{\cal H}e}{2}(\eta^2 + r^2)$ (in Reg. I)\\

& & $R= {\cal H}e \eta r$ (in Reg. I)\\

 & $x=\frac{e^{\alpha\xi_1}}{\alpha} \cosh(\alpha\xi_0)$ & $T= {\cal H}e \eta r$ (in Reg. II)\\
 
 & & $R= \frac{{\cal H}e}{2}(\eta^2 + r^2)$ (in Reg. II)\\
 &&\\
 
{ iii) Coordinate span} & iii) $-\infty \le t \le \infty$; $0\le r \le \infty$ & iii) $0 \le \eta \le \infty$; $0\le r \le \infty$ \\
 & $- \infty \le (\xi_0, \xi_1) \le \infty$ & $0 \le T \le \infty$; $0\le R \le \infty$ \\

 &&\\

 { iv) Completeness} & iv) $(t,r)$ complete  & iv) Both are complete \\
 & but  $(\xi_0, \xi_1)$ incomplete & \\
 
 &&\\
 
{v) Overlaps} & v) ($\xi_0, \xi_1$) overlap  & v) overlap in entire spacetime \\

 & only in the quarter of ($t,r$) &  \\
 
 &&\\
 
{vi) Observers} & vi) Rindler & vi) Cosmological \\
&&\\

{ vii) Spaceslices} &  vii) $\xi_1=const.$   & vii) $r=const.$ \\
 & hyperbola $x^2 - t^2 = \frac{e^{2\alpha\xi_1}}{\alpha^2}$,  & $T = \frac{R^2}{2 {\cal H} e r^2} + \frac{1}{2}{\cal H} e r^2$ (region-I)\\
 & (for $x\ge \mid{t}\mid$) & $R = \frac{T^2}{2 {\cal H} e r^2} + \frac{1}{2}{\cal H} e r^2$ (region-II) \\
 &  & (joined at Hubble scale)\\
 
 &&\\
 
{ viii) Timeslices} & viii) $\xi_0=const.$ & viii) $\eta=const.$  \\
 & $ \xi_0  = \frac{1}{\alpha}\tanh^{-1}{(t/r)}$ & $T = \frac{R^2}{2 {\cal H} e \eta^2} + \frac{1}{2}{\cal H} e \eta^2$ (region-I) \\ 
&  &  $R = \frac{T^2}{2 {\cal H} e \eta^2} + \frac{1}{2}{\cal H} e \eta^2$ (region-II)\\ 
  & &\\ \hline

\multicolumn{3}{c}{\begin{normalsize} Table 1: Analogy between the Rindler-Minkowski and FRW-New coordinates.\end{normalsize}} \\
%\caption[]{aa}
\end{tabular}
\label{table1}
\end{center}
  
Now, we want to compare the studies based on the quantum field theory side which will give a side by side picture of the Unruh effect with the present study. This is also shown in the following tabular form:
 
\begin{center}
\begin{tabular}{|c| l | l |} \hline
 Item  & Rindler -- Minkowski & FRW -- New coord. \\ \hline
 & \begin{footnotesize}\end{footnotesize} & \begin{footnotesize}\end{footnotesize} \\
{ i) Vacuum states} & i) Rindler vacuum: $|0_{R}\rangle$  & i) Cosmological vacuum: $|0_{C}\rangle$ \\
 & Minkowski vacuum: $|0_{M}\rangle$   &  $T$-vacuum: $|0_{T}\rangle$ \\
 & &\\
{ ii) Physical observers} & ii) Both Rindler and Minkowski & ii) Cosmological\\
&&\\

{ iii) Physical vacuum} & iii) $|0_{M}\rangle$ & iii) Both $|0_{C}\rangle$ and $|0_{T}\rangle$\\
&&\\
 %{ iv) Bogolyubov coeffcient} & iv) $|\beta_{\omega\Omega}|^2 = \frac{\Omega e^{-\frac{\pi\Omega}{a}}}{4\pi^2 a^2 \omega}  \Gamma(-i\Omega/a) \Gamma(i\Omega/a)$ & iv) $|\beta_{\omega\Omega}|^2 = \frac{\omega }{8 e {\cal H} \pi^2 \Omega^2 } \left(1+ \sin(\frac{\omega^2}{\Omega e {\cal H}})\right)\Gamma^2[\frac{1}{2}, \frac{\omega^2} {2 \Omega e {\cal H}}]$\\
{v) Number density } & v) $\langle n_{\omega} \rangle = \frac{1}{e^{2\pi\omega/\alpha} -1}$  & v) Numerically plotted \\
& &  in Fig. \ref{fig2} \\
&&\\
{vi) Field modes} & vi) Complete in Minkowski & vi) Complete in both  \\
& but incomplete in Rindler & frames\\
&&\\
{ vii) Components of } &  vii) $\lim_{u \rightarrow 0}\langle 0_{R}| T_{uu} |0_{R}\rangle \rightarrow \infty$   & vii) $\lim_{u\rightarrow 0} \langle 0_{T}| T_{uu} |0_{T}\rangle = 0$ \\
REMT& $~~~~~\lim_{v\rightarrow 0}\langle 0_{R}| T_{vv} |0_{R}\rangle \rightarrow \infty$& $~~~~\lim_{v\rightarrow 0} \langle 0_{T}| T_{vv} |0_{T}\rangle = 0$ \\
  & &\\ \hline
\multicolumn{3}{c}{\begin{footnotesize} Table 2: Comparison between the particle creation in this paper with the Unruh effect.\end{footnotesize}} \\%While in Unruh effect both Minkowski  and  \end{footnotesize}} \\
%\multicolumn{3}{c}{\begin{footnotesize}  Rindler observers are of physical interest, here only the cosmological observer has a physical interest. However, while in Unruh  \end{footnotesize}}\\
%\multicolumn{3}{c}{\begin{footnotesize}effect only Minkowski vacuum is physical (and Rindler is not),  here both the cosmological vacuum and the $T$-vacuum are physical.\end{footnotesize}} \\
\end{tabular}
\end{center}

\section{Conclusions}\label{s7}
To conclude, we have added a new example for particle creation in the cosmological background -- specifically, in the radiation dominated early universe. It is an extension of our study \cite{modak1} which further clarifies various aspects of quantum field theory in the radiation dominated stage of early universe. Particularly, we have studied the cosmological observers' view of the universe who have a peculiar motion in the frame of the new coordinates ($(T,~R)$ coordinates) and behave like the accelerated observers in the flat space. Our motivation for focussing on those observers is rooted in the fact that in cosmology we have a tendency to understand the universe with respect to comoving observers who are closely related with the cosmological observers just by a redefinition of the time coordinate. Also, due to conformal flatness in the cosmological observers frame it is relatively clear to study field theory in their frame. We, in the Earth, do not share the cosmological observer's frame, but it is widely accepted that by subtracting all relative motions, such as (a) the Earth's rotation around the Sun, (b) the Sun's motion relative to the Local Standard of Rest (LSR), (c) the motion of LSR orbit in the Milky Way, (d) The Milky Ways's motion relative to the Local Group (LG), (e) the LG's infall in Virgo Cluster of galaxies and finally (f) the speeding of Virgo cluster towards ``The Great Attractor'', we can get cosmological observer's view of the Universe. In this work, apart from their peculiar trajectories in $(T,~R)$ frame, these observers are shown to be exposed to a radiation due to a new gravitational particle creation process. This particle creation process is a new discovery in this paper and it is possible due to the existence of a new vacuum state (we refer to that as $T$-vacuum) that appears to be a particle excited state  in  cosmological observer's reference frame. 

We also made an analysis to show that the newly found $T$-vacuum is a physical vacuum state since it does not attribute undesired divergences in any component of the renormalized energy momentum tensor (REMT). We calculated REMT in $T$-vacuum both for the cosmological observers and later for comoving observers. The result shows that in both frames (cosmological observer's and comoving observer's frame) the energy density is negative which is effectively giving an effect of repulsive gravity and giving an extra push to the expansion rate. Also, this quantum effect give a non-zero pressure and momentum densities which are otherwise non-existing for the radiation that is driving the background spacetime. We gave a detailed comparison of this effect with the Unruh effect.
 We feel that understanding this effect with finer details is important which includes a  generalization to the  realistic four dimensional set up. 
 
One may, keeping the four dimensional set up in the hindsight, question the motivation behind the $T$-vacuum state, from a phenomenological point of view. In fact, asking if the massless scalar field in the radiation dominated phase did occupy the state of $T$-vacuum  is very interesting. Even if one can prove the mathematical well-defined ness of this quantum state it is not guaranteed that it has to be an actual state of the scalar field at any point of time. In fact, seeking a valid answer to this question could lead us to the following novel perspective.

First we note that, if a quantum state is mathematically well-defined, we cannot ignore it a-priory even if we do not have a physical understanding behind it. As the next cross-check, we should always look out for observational/experimental evidences related with the said quantum state to test its physical validity. 

In this particular context, we first realize that the observed universe, even at the time of CMB decoupling, is almost isotropic and homogeneous. On the other hand, the $T$-vacuum is defined here in an inhomogeneous background and it is expected to reflect the symmetry of the background metric, even in 4 dimensions. {\it Note that, if a metric is inhomogeneous it is necessarily anisotropic}. That imply, in turn, {\it $T$-vacuum is anisotropic in nature, even in 4 dimensions}. Can we rule out such a state from the observation? The answer is No! In fact, it is on the contrary. Notice that, by observation, we can rule out the physical validity of the $T$-vacuum only on the ground of non-detection of any anisotropy whatsoever. But this is not the case! There is anisotropy in CMB, however small it is. The quantum origin of these anisotropies is still not fully understood \cite{sud}. This brings us to the question, could the $T$-vacuum be able to contribute to such anisotropy in the CMB? We do not know the answer for sure, but we do want to point out following facts which is encouraging and make a case in favor of $T$-vacuum:

\begin{itemize}
\item $T$-vacuum is defined before the CMB map is observed. CMB photons get decoupled only when matter started dominating over the radiation. That is, in the beginning of matter dominated phase. Before this, the universe is not transparent because photons do not move freely, and they are scattered by atoms. Therefore, an anisotropic vacuum state, such as $T$-vacuum, cannot be ruled out from the CMB observation which captures the beginning of matter domination and not the phase of the universe we are considering here.

\item In fact, $T$-vacuum may help us to give an alternative explanation of the anisotropy (full or a part of it) observed in the CMB map. We want to point out here that the quantum origin of (classical) temperature anisotropy in CMB is a highly controversial topic among quantum cosmologists which has also gained attention from the community working on the foundation of quantum theory. The controversy is rooted in the use of Bunch-Davies (BD) vacuum state, defined in the early inflationary universe, which is usually used to explain the classical temperature anisotropies in the sky. This effort has landed into a deep controversy. The main reason of this criticism to the standard approach which uses the BD vacuum is the following -- since the symmetry of the BD vacuum is homogenous and isotropic, and since there cannot be a unitary evolution of quantum dynamics which can break the symmetry of this initial state, it is very hard to explain, quantum mechanically, the origin of the classical anisotropies of temperature field \cite{sud}. 

On the other hand, $T$-vacuum is necessarily anisotropic, and then total radiation due to the particle creation that would be accumulated during the entire radiation dominated phase could contribute to the anisotropic part of the CMB radiation.

This is not a quantitative claim but at the phenomenological level we see the motivation behind the $T$-vacuum state. CMB anisotropies, thus, could give a test for the $T$-vacuum and we would know if the massless scalar field was ever caught in the state of $T$-vacuum. This paper is not a stage for such a query but we hope to consider it in future.

\end{itemize}

%%%%%%%%%%%%%%%%%%%%%%%%%%%%%%%%%%%%%%%%%%%
\section{Acknowledgements}
I thank Professor Thanu Padmanabhan for several discussions which led me to this project and for his key inputs in the early stages of the work. I also thank Professor Satoshi Iso for his key conceptual and technical inputs during several discussions. Parts of my research was carried out while I was visiting KEK, Japan and California State University, Fresno. This research is supported by the CONACyT Project No. CB17- 18/A1-S-33440 (Mexico).
 
%%%%%%%%%%%%%%%%%%%%%%%%%%%%%%%%%%%%%%%%%%%

\end{document}